\documentclass[preprint]{emulateapj}
\shorttitle{Starlight Demonstration of Dragonfly}
\shortauthors{Nemanja Jovanovic}

\usepackage{graphicx}
\usepackage{float}
\usepackage{booktabs}
\usepackage{color}
\usepackage[mathscr]{eucal}

\begin{document}

\title{
  Starlight Demonstration of the Dragonfly Instrument: an Integrated Photonic Pupil Remapping Interferometer for High Contrast Imaging
}

\author{
  N.~Jovanovic\altaffilmark{1,2,3}, P. G.~Tuthill\altaffilmark{4,5}, B.~Norris\altaffilmark{4}, S.~Gross,\altaffilmark{1,5}, P.~Stewart\altaffilmark{4}, N.~Charles,\altaffilmark{4}, S.~Lacour\altaffilmark{6}, M.~Ams\altaffilmark{1,5}, \\ J. S. Lawrence\altaffilmark{1,2,3}, A. Lehmann\altaffilmark{1,2}, C. Niel\altaffilmark{1,2}, J. G. Robertson\altaffilmark{4}, G. D. Marshall\altaffilmark{1,5}, M. Ireland\altaffilmark{1,2,3}, A. Fuerbach\altaffilmark{1,5} and M. J. Withford\altaffilmark{1,2,5}}

\altaffiltext{1}{MQ Photonics Research Centre, Dept. of Physics and Astronomy, Macquarie University, NSW 2109, Australia}
\altaffiltext{2}{Macquarie University Research Centre in Astronomy, Astrophysics and Astrophotonics, Dept. of Physics and Astronomy, \\ Macquarie University, NSW 2109, Australia}
\altaffiltext{3}{Australian Astronomical Observatory (AAO), PO Box 296, Epping NSW 1710, Australia}
\altaffiltext{4}{Sydney Institute for Astronomy (SIfA), Institute for Photonics and Optical Science (IPOS), School of Physics, \\University of Sydney, NSW 2006, Australia}
\altaffiltext{5}{Centre for Ultrahigh Bandwidth Devices for Optical Systems (CUDOS), Sydney, NSW, Australia}
\altaffiltext{6}{Observatoire de Paris, 5 place Jules Janssen, Meudon, France}

\email{jovanovic.nem@gmail.com}

\begin{abstract}
In the two decades since the first extra-solar planet was discovered, the detection and characterization of extra-solar planets has become one of the key endeavors in all of modern science. Recently Òdirect detectionÓ techniques such as interferometry or coronography have received growing attention because they reveal the population of exoplanets inaccessible to Doppler or transit techniques, and moreover they allow the faint signal from the planet itself to be investigated. Next-generation stellar interferometers are increasingly incorporating photonic technologies due to the increase in fidelity of the data generated. Here, we report the design, construction and commissioning of a new high contrast imager; the integrated pupil-remapping interferometer; an instrument we expect will find application in the detection of young faint companions in the nearest star-forming regions. The laboratory characterisation of the instrument demonstrated high visibility fringes on all interferometer baselines in addition to stable closure phase signals. We also report the first successful on-sky experiments with the prototype instrument at the $3.9$-m Anglo-Australian Telescope. Performance metrics recovered were consistent with ideal device behaviour after accounting for expected levels of decoherence and signal loss from the uncompensated seeing. The prospect of complete Fourier-coverage coupled with the current performance metrics means that this photonically-enhanced instrument is well positioned to contribute to the science of high contrast companions.
\end{abstract}

\section{Introduction}
The overwhelming majority of the $800$ or so extra-solar planets confirmed to date\footnote{http://exoplanet.eu} have been detected using the well-established transit~\citep{Burocki2010} or Doppler~\citep{Tinney2002} techniques. Although these ÒindirectÓ detection techniques are prolific in detection yield, both are strongly biased to detecting large planets in close proximity to their parent star. Such selection biases have delivered a highly skewed picture of the mass/size/orbit distribution of extra-solar planets which limits our understanding of planetary evolution and leaves formation models unconstrained~\citep{Marois2010}. Unlike most indirect methods, high contrast imaging techniques encompassing both coronography~\citep{Guyon2003} and interferometry~\citep{Angel1997,Tuthill2000} make direct use of the light from the companion itself for scientific investigation. As a result exoplanetary companions can be studied in any orbit/period~\citep{Serabyn2010} down to and even below an angular separation corresponding to the diffraction limit ($\approx\lambda/D$ where $\lambda$ is the wavelength of light and $D$ the diameter of the telescope mirror)~\citep{Huelamo2011}. In addition, it is possible to study planetary systems with arbitrary orbital inclination to the observer, giving access to a larger population of potential targets. However, it should be noted that these techniques are currently limited to observations of young massive and/or actively accreting objects (warm and bright) while indirect methods can be used to detect older planets (cooler and hence fainter).

One powerful interferometric technique is aperture masking~\citep{Tuthill2000} which offers precise calibration of wavefront structure for angular separations from the target star of around $\lambda/D$. This is achieved by recording Fizeau interferograms generated by a non-redundant sparse-aperture mask placed in a re-imaged telescope pupil-plane. Such a scheme has been shown to be highly robust against the degrading effects of atmospheric phase-noise, particularly when the self-calibrating closure phase observable~\citep{Baldwin1986} is utilised~\citep{Lloyd2006,Lacour2011}. This technique has been responsible for a series of recent high profile discoveries~\citep{Kraus2008,Kraus2011}, including the first two newborn substellar companions captured in the process of formation~\citep{Huelamo2011,Kraus2012}. Such young systems are valuable as they offer insights into planetary formation and allow us to constrain planetary evolution models. Despite these successes, aperture masking has several limitations including low throughputs (typically $\approx5$--$10\%$ of the stellar photons are transmitted by a mask), incomplete Fourier coverage, leakage of residual atmospheric noise due to phase corrugations across each sub-aperture and the two-dimensional nature of the interferograms makes polychromatic setups involving cross-dispersion difficult to realise optically.

\section{Harnessing Photonics}
Photonic technologies promise dramatic gains in the precision with which an interferometer can be fabricated, the fidelity of the data recovered, as well as in the robust, stable performance over a wide range of challenging environments. A recent photonic-reformulation of aperture masking replaces the mask with an optical system which feeds a number of single-mode optical fibres that sample the pupil-plane~\citep{Chang1998,Perrin2006,Lacour2007,Tuthill2010}; a scheme that addresses many of the current shortcomings. Firstly, because the light within a single-mode guide propagates with planar wavefronts, an interference pattern formed between the beams of two such guides will have a greater signal-to-noise ratio of fringe visibility (contrast) than if the atmospherically corrupted wavefronts were combined ($\approx100\times$ increase was demonstrated for the FLUOR instrument~\citep{Coude1994}). Secondly, fibres can be routed from a $2$-dimensional pupil-plane to any configuration required at the output; an operation referred to as pupil remapping. This allows for the fibres to be reformatted for non-redundant beam recombination at the output while sampling the entire (highly redundant) telescope pupil at the input, enabling all available starlight to be used (up to $20\times$ increase in throughput compared to aperture masking if light is injected with a high-Strehl ratio adaptive optics (AO) system). A filled entrance pupil also yields complete Fourier sampling; a dramatic gain over a sparsely sampled mask. By reformatting waveguides into a linear output array, starlight can be injected into an integrated-photonic beam-combiner~\citep{Berger2001,Laurent2002,Benisty2009}, and/or cross-dispersed for the extraction of simultaneous spectral and spatial information. A key advantage of conducting beam combination on-chip is the possibility for nulling~\citep{Labadie2007}, whereby the signal from the bright stellar source is cancelled out in one output port so that the faint signal from a companion can be detected~\citep{Bracewell1978}.

The challenge with any optical stellar interferometer, including a pupil-remapper, is that the paths through the instrument must be matched to within the coherence length of the light ($L_{coh.}\approx50~\mu$m is typical for the bandwidths employed in the near-infrared). This is difficult to achieve with optical fibres because environmental influences such as temperature and strain must be carefully controlled over lengths up to several meters. Indeed, an optical fibre based pupil remapper known as the FIRST instrument has been demonstrated, but maintaining stable performance has proved challenging \citep{Huby2012,Kotani2010}. An ideal alternative is to use integrated-photonic components where all the waveguides can be embedded in a single miniaturized monolithic chip. In this way all the paths experience a near-identical environment, with precision control at fabrication allowing for tailoring of the trajectory of each guide in order to achieve path-length matching. A promising technology for achieving the desired $3$-dimensional waveguide structures required is the laser direct-write technique~\citep{Nolte2003,Thomson2011}. In essence this involves sculpting tracks of modified material, which act as waveguides, within the bulk of a dielectric (e.g. fused silica) with the use of a tightly focused femtosecond-pulsed laser~\citep{Gattass2008}.

Here we present an integrated pupil-remapping interferometer (IPRI) known as Dragonfly, for high contrast imaging. This new instrument concept is ultimately intended to target ambitious scientific goals in exoplanetary science. The instrument relies on the unique synergy between several micro-scale optical technologies including the first photonic path-length-matched pupil-remapper, a steerable segmented-mirror array and micro-lens arrays. This paper presents the conceptual and practical design features of the instrument, and summarises the results from the laboratory tests and the successful commissioning at the Anglo-Australian Telescope in section~\ref{s:exp}. From these results we extrapolate a predicted on-sky performance for the instrument when used in tandem with an AO system in section~\ref{s:pred}. In section~\ref{s:space}, we round-out this body of work by outlining the narrow but critical parameter space at small angular separations from the parent star, where the instrument will have few competitors and be able to make a key contribution to the field of exoplanetary science.

\begin{figure*}
\centering 
\includegraphics[width=0.99\linewidth]{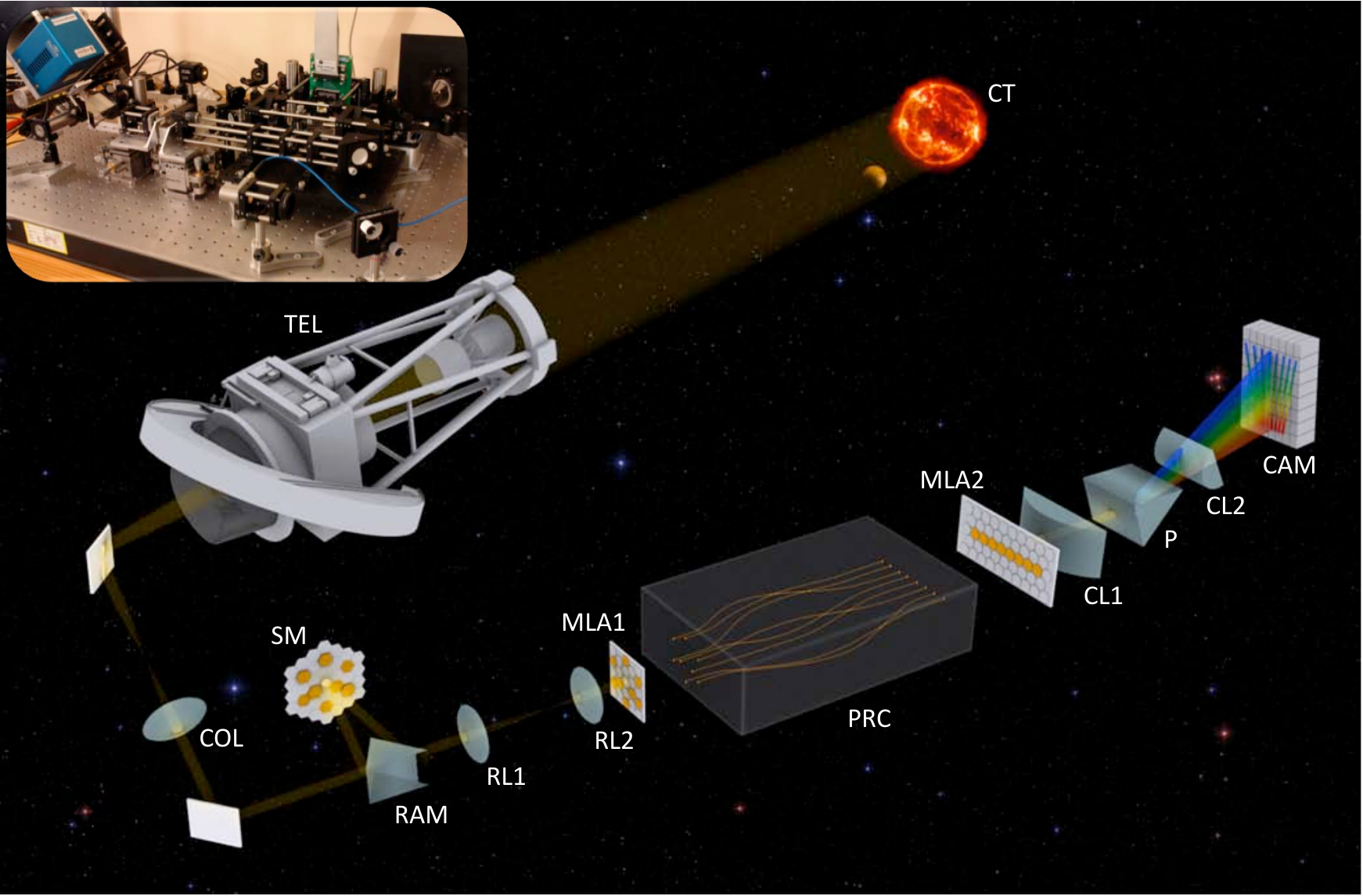}
\caption{Schematic diagram of the IPRI. Light from the celestial target (depicted as a star with an orbiting planet in this figure) is collected by the telescope before propagating through the instrument and finally being collected at the detector. Orange circles indicate the correspondence between the segments used on each optical component. The diagram is figurative only: the aspect ratio and scale are not preserved. To follow the beam path downstream we have: CT - celestial target, TEL - telescope, COL - collimating lens, RAM~-~right angled mirror, SM - segmented mirror (IrisAO - PTT$111$), RL$1$ - relay lens $1$ ($200$~mm focal length achromat), RL$2$ - relay lens $2$ ($10$~mm focal length aspheric), MLA$1$ - micro-lens array $1$ ($30~\mu$m pitch), PRC - pupil remapping chip, MLA$2$ - micro-lens array $2$ ($250~\mu$m pitch), CL$1$ - cylindrical lens $1$ ($200$~mm focal length), CL$2$ - cylindrical lens $2$ ($20$~mm focal length), P - prism, CAM - camera (Xenics-Xeva-$1.7$-$640$). Inset: Photograph of the IPRI instrument on a $900\times600$~mm breadboard.}
\label{fig:dragonfly}
\end{figure*}

\section{Experiments}\label{s:exp}
An artist's impression of the IPRI is shown in Figure~\ref{fig:dragonfly}. Starlight from a celestial target is collected by the telescope and routed to the photonic pupil-remapping chip into which it is injected by a micro-lens array (hexagonal lattice, $30~\mu$m pitch). Although there were no specific limits to the spectral content of the signal that was coupled into the remapper, the IPRI was optimised for operation across the H-band. Each micro-lens of the array had a one-to-one correspondence with each of the $37$ elements of a segmented mirror upstream and waveguides of the chip downstream. Since the segmented mirror and the micro-lens array were positioned in the conjugate pupil planes of the telescope respectively, the tip/tilt functionality of each mirror segment was used to steer the sub-pupil beams carefully in order to fine-tune the coupling into each waveguide. The mirror could also be used to compensate for path-length mismatches via its piston functionality. Although this was not utilised in the on-sky observations, it proved a very valuable feature for our laboratory testing campaign. To minimise the amount of stray light injected into the pupil remapper a mask was placed in close proximity to the front of the segmented mirror (i.e. close to the pupil plane, not shown in Fig.~\ref{fig:dragonfly}). The mask was tilted at an angle to reject light from the segments that were not used and had circular sub-apertures ($90\%$ of the size of a mirror segments) to transmit the light for the segments that were used. After passing through the waveguides, a second micro-lens array (hexagonal lattice, $250~\mu$m pitch) was used to re-collimate the emergent light before it was cross-dispersed by means of a $60^{\circ}$ Flint glass prism and focused with anamorphic optics onto a research grade InGaAs camera ($20~\mu$m pixels). The linear dispersion was $29$~nm/pixel (at $1.55~\mu$m) giving a resolving power ($R=\lambda/\Delta \lambda$ where $\Delta\lambda$ is the bandwidth of the light), per pixel of $53$.

\subsection{The Pupil-Remapper Prototype}
\subsubsection{Design and Fabrication}
\begin{figure}
\centering 
\includegraphics[width=0.99\linewidth]{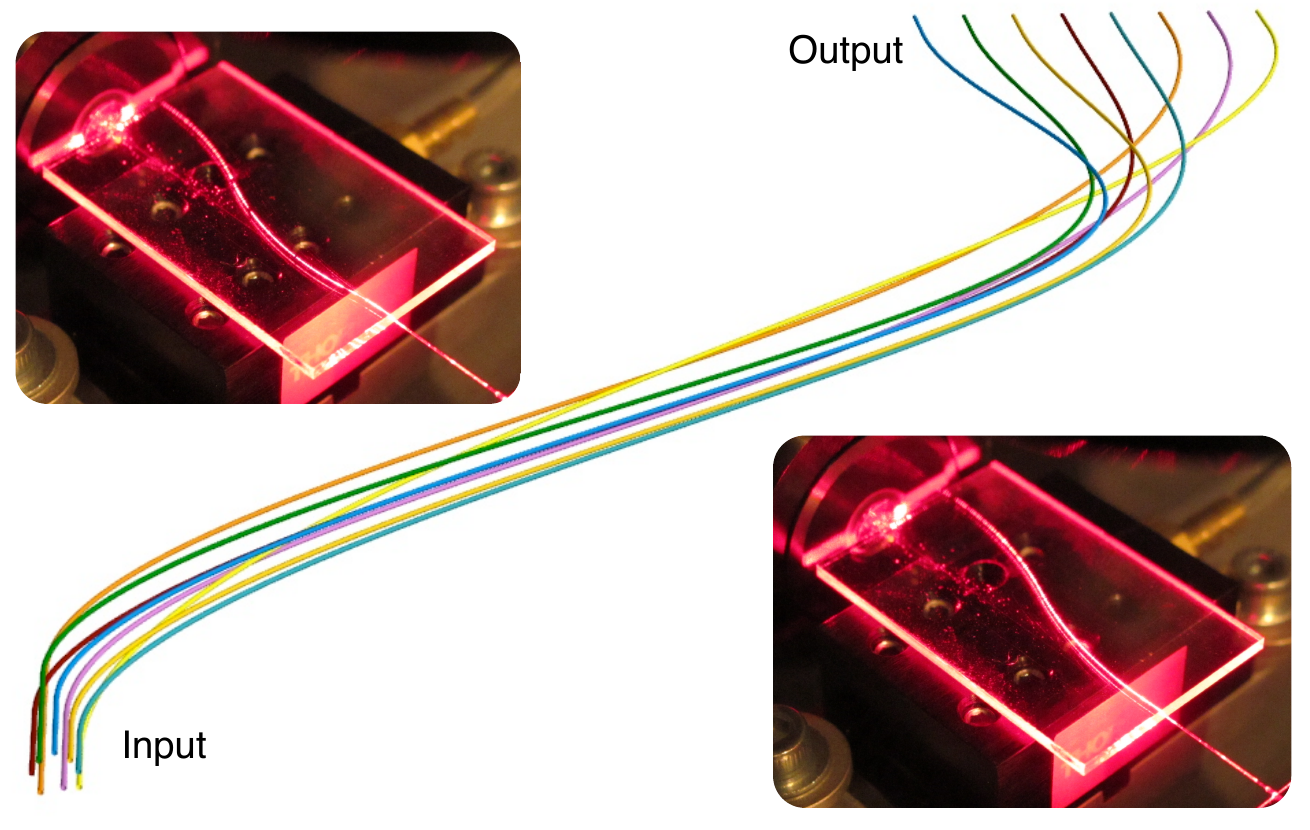}
\caption{A CAD rendering of the 3D paths taken by the 8-waveguide pupil remapper. Note: the aspect ratio for the drawing greatly inflates the transverse coordinates to exaggerate the bends and curves in the guides. Insets show images of two of the waveguides illuminated with red laser light.}
\label{fig:pupilremapper}
\end{figure}

A prototype pupil-remapping chip was designed and is depicted in Figure~\ref{fig:pupilremapper}. It consisted of $8$ waveguides which were constructed by interpolating with cubic spline functions between nodal points placed along each path. By moving the nodal points it was possible to attain physical path length matching between all waveguides to within $100$~nm while maintaining a minimum separation between guides of $30~\mu$m in order to minimise cross-talk~\citep{Charles2012}. This design process only matched the path lengths physically because the optical contributions, such as the reduced effective index that a mode experiences as it propagates around a bend, are minor contributors that did not increase the path-length mismatch beyond the coherence length of the light. The design proved to be extremely challenging and the resulting device, is to the best of our knowledge the first $3$D, path-length matched chip consisting of unique routes (i.e. each waveguide is unique and not simply a mirror of any other).

The device was inscribed into a boro-aluminosilicate glass substrate (Corning Eagle$2000$) with dimensions of $30\times20\times1.1$ mm ($L\times W\times H$) by the laser direct-write technique. An ultrafast titanium sapphire oscillator (Femtolasers GmbH, FEMTOSOURCE XL $500$, $800$~nm centre wavelength, $<50$~fs pulse duration) with a $5.1$~MHz repetition rate, was used for inscription. The laser was focused into the sample using a $100\times$ oil immersion objective lens (Zeiss N-Achroplan, \textit{NA}$=1.25$, working distance of $450~\mu$m). In order to prevent the immersion oil boiling during writing, the average power of the laser was controlled by using an external electro-optic pulse picker (BME KG) to reduce the repetition rate to $1.28$~MHz. Pulse energies of $160$~nJ were used in conjunction with a translation velocity of $8.3$~mm/s in order to create waveguides that supported a single-guided mode at $1.55~\mu$m. A set of Aerotech, air-bearing translation stages were used to smoothly translate the sample in $3$ dimensions. The entire device was written within half a minute. The device was then ground and polished to reveal the waveguide ends. As a result of the high repetition rates used all waveguides presented herein were written in the cumulative heating regime \citep{Eaton2005}.

\begin{figure}
\centering 
\includegraphics[width=0.90\linewidth]{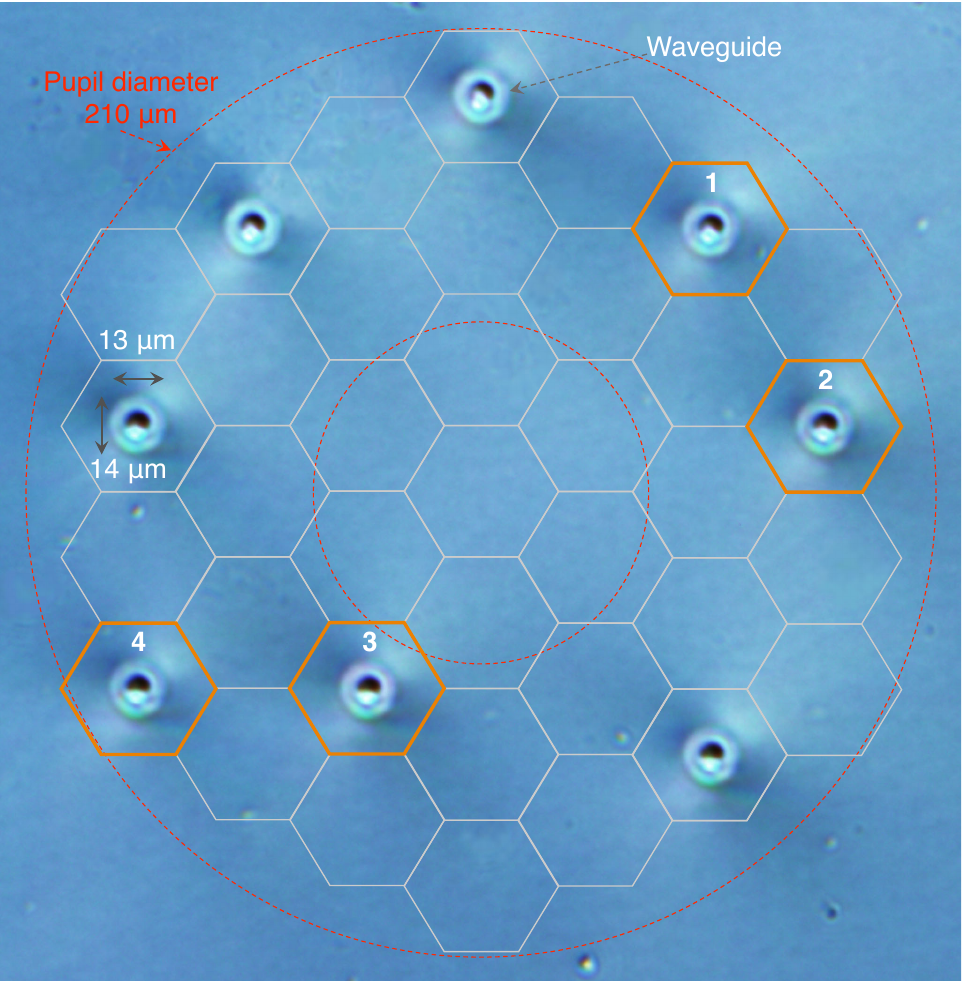}
\caption{Cross-sectional micrograph of the input facet of the pupil-remapping chip. Eight waveguides are clearly visible each of which consists of a complex index profile made up of a high index central region (bright central spot, positive index contrast $>6\times10^{-3}$) bordered by a low index region (dark spot, negative index contrast of $<-1\times10^{-3}$) surrounded by a high index ring (weak positive index contrast $<3\times10^{-3}$). The optical field at $1.55~\mu$m is guided over the entire index profile. The image is overlaid with a hexagonal grid depicting the overlap with the elements of the segmented mirror and the injection micro-lens array. The dashed red overlay shows the projection of the$3.9$~m primary telescope mirror (the pupil).}
\label{fig:micrograph}
\end{figure}

The $8$ single-mode waveguides sampled the re-imaged 2D conjugate pupil plane of the telescope which had a diameter of $210~\mu$m at the Coude focus (see Figure~\ref{fig:micrograph}). The starlight was reformatted into a $1$D equidistantly spaced ($250~\mu$m) output array that spanned a total of $1.75$~mm. In Figure~\ref{fig:pupilremapper} it can be seen that a lateral side step was used. This was done in order to prevent unguided stray light reaching the output face where the waveguides terminate which would induce phase errors in the measurement~\citep{Norris2011}. The distribution of the waveguides at the input of the remapper (Figure~\ref{fig:micrograph}) was chosen to conform to the hexagonal lattice of the elements of both the segmented mirror and the micro-lens array (overlaid in Figure~\ref{fig:micrograph}). A projection of the pupil of the primary mirror has also been overlaid in Figure~\ref{fig:micrograph}. The inscribed circle of each hexagonal segmented mirror element and its corresponding micro-lens, subtended an area with $56$~cm diameter on the primary mirror. For this first prototype a subset of $8$ of the available lenslets was used in order to facilitate the characterization of the instrument, with the specific pattern chosen to provide adequate Fourier sampling. However, utilizing all available lenslets is possible (with some work) and will be the focus of future prototypes and indeed the final instrument.

\subsubsection{Characterisation}
\begin{figure}
\centering 
\includegraphics[width=0.85\linewidth]{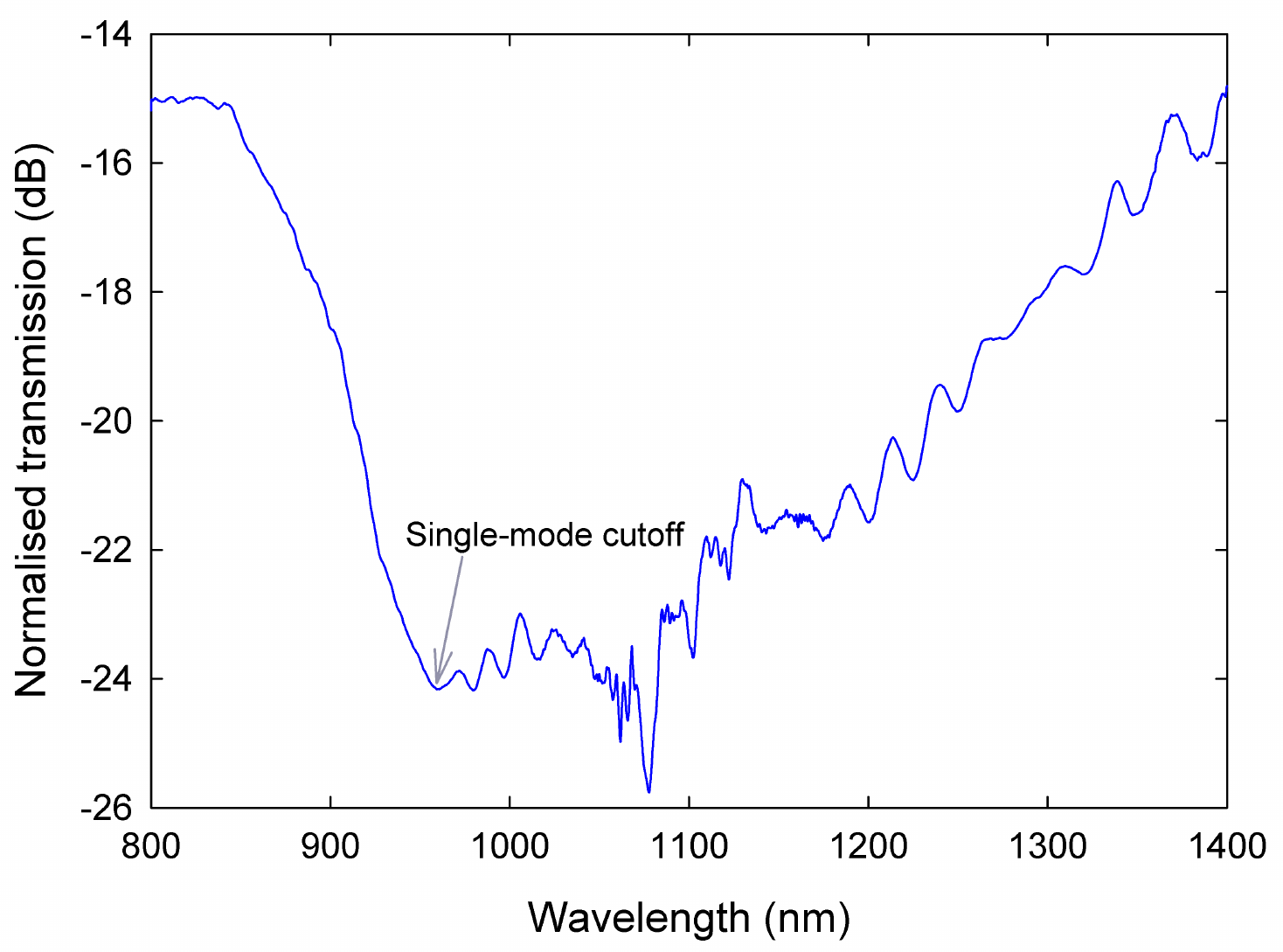}
\caption{Normalised broadband transmission spectrum for one of the waveguides in the pupil-remapping prototype. The single-mode cutoff wavelength is the point where the transmission recovers abruptly which has been highlighted in the Figure. The sharp features in the spectrum between $1.05$ and $1.10~\mu$m are due to the pump laser and the associated Raman lines of the super continuum source which were not entirely normalised.}
\label{fig:shortlcutoff}
\end{figure}

The prototype pupil-remapper was designed to operate around $1.55~\mu$m in the astronomical H-band. The slightly elliptical waveguides ($13\times 14\pm1 ~\mu$m) are depicted in Figure~\ref{fig:micrograph} and consisted of non-step index profiles with peak index contrasts of $>6\times10^{-3}$. Before the waveguides could be used for broadband interferometric applications, it was important to first determine the extent of the single-mode regime which is where spatial filtering could be exploited, as described above. The single-mode cutoff of the waveguides was determined by the \textit{transmitted power method} \citep{Lang1994}. This involves exciting the modes of the structure and investigating the broadband transmission characteristics to identify the point in the spectrum where the losses drop abruptly (i.e. the transmission recovers). This point signifies that a second mode is now bound to the guide and can carry energy with relatively low loss which at longer wavelengths it was radiating away. The light from a broadband (super continuum) light source was injected and collected from the waveguides by butt coupling the device to single-mode optical fibres (SMF-28). It should be noted that the fibres were not aligned for optimal transmission but rather they were offset laterally by $6~\mu$m in order to excite the higher order mode of the guide. The transmitted spectrum was recorded by an optical spectrum analyzer (OSA). The spectrum was normalised to that which was transmitted across a butt couple between the probe and collection fibres without the chip in between. A normalised transmission spectrum for one of the waveguides from the remapper is shown in Figure~\ref{fig:shortlcutoff} with the cutoff wavelength highlighted as the point where there was a dramatic recovery in the transmission. The mean single-mode cutoff wavelength for the 8 waveguides was $0.95\pm0.03~\mu$m which enabled the IPRI to operate in a highly broadband fashion across the entire Y, J and H bands ($0.96$--$1.80~\mu$m). It should be made clear that the high losses shown in Figure~\ref{fig:shortlcutoff} are not indicative of the waveguides performance but are the result of the off-axis injection of the probe light in order to excite the higher order modes required for the measurement of the single-mode cutoff. 

The throughput of the pupil-remapper was characterised using the same fibre optic circuit used for the single-mode cutoff measurements, with a few slight modifications; the light source was exchanged for a narrow linewidth laser diode operating at $1.55\mu$m and a Ge-photodetector and power meter were used in place of the OSA. High precision, flexure translation stages were used to align the fibres with the waveguides, one at a time. The maximum transmitted power was recorded and normalised with respect to the power transmitted across a butt couple between the probe and collection fibres without the chip in between. Index matching immersion oil was used at all interfaces to remove Fresnel reflections. The extent of cross-talk between the waveguides was examined by fixing the probe fibre for maximum alignment into a particular guide while the collection fibre was scanned between each of the waveguides at the output. The throughputs ranged from $5$--$50\%$ for the $8$ waveguides. The components of the loss are summarised in Table~\ref{tab:loss}. It can be seen that $\approx20\%$ is due to the non-negligible absorption by the Eagle$2000$ substrate~\citep{Jovanovic2012}, $5\%$ is due to a mode mismatch between the fibre and the waveguide modes (which was determined by calculating the overlap integral of the two fields), $30$--$93\%$ is attributed to bend and transition losses, and $<0.1\%$ is the result of cross-coupling. It is important to point out that the coupling loss reported was for the case when optical fibres were used to inject into the pupil-remapper, which is not the optical configuration for the IPRI. To account for this, the mean coupling efficiency was established with the micro-lens array to be $60\pm2\%$. It should be noted that when the pupil-remapper was used in conjunction with micro-lenses, there was an additional Fresnel reflection loss at each facet of the pupil-remapping chip as antireflection (AR) coatings were not employed in this body of work. The higher than anticipated bend losses are the result of both the overly tight bend radii that were used in the design of the prototype and the subsequent manufacturing errors that this caused when writing at high speeds.

\begin{table}
\caption{Breakdown of losses in the current pupil-remapping prototype and a prediction of the performance of an optimised device. The coupling losses shown are the same for all waveguides/lenses, the bend/transition losses represent the range of those for the $8$ waveguides while the cross-coupling loss shows the maximum value measured amongst the $8$ guides. The total throughput of the current pupil-remapper as used within the IPRI is summarised in the last line of the table along with a forecast for an optimised system.}
\label{tab:loss}
\begin{tabular}{lcc}
\toprule
Component of loss 					& 	Currently 				& 		Predicted minimum 		\\
								&	measured				&		achievable			\\
								& 	losses (\%)			&     		 losses (\%) 		\\ \toprule
Absorption by substrate  				&  	$20\pm1$ 			& 		$<0.5$ 		\\
Coupling losses 					&  						&   					\\
\hspace{3 mm} Using optical fibres		&      $5\pm1$ 				& 					\\
\hspace{3mm} Using micro-lens array	&      $40\pm2$ 			& 		$20$ 		\\
Bend/transition losses  				&     $30$--$93$ 			& 		$<15$ 		\\
Cross-coupling 					&      $<0.1$				&		$<0.05$ 		\\  \midrule
Total throughput of					&     						&					\\
remapper within IPRI 		 		&	$3-31$				&		$>68$ 		\\ \bottomrule
\end{tabular}
\end{table}

Although this level of performance was adequate for initial on-sky tests presented below, we believe we can minimise the losses of the pupil-remapper in future. This can be achieved by increasing the index contrast of the guides in order to optimise the coupling with the micro-lens array ($80\%$), and minimise the cross-talk ($<0.05\%$). It should be made clear that the predicted maximum level of coupling of $80\%$ is based on the limit when we consider the overlap between the Airy-shaped focal spot and the Gaussian mode of the waveguide~\citep{Shaklan1988}. In addition by utilising circular arcs instead of cubic spline functions and subsequently maximising the bend radii, it should be possible to reduce the bend losses ($<15\%$) provided that waveguides are written at lower translation speeds to overcome manufacturing difficulties. By using a highly transparent substrate like fused silica for example, the absorption losses in the substrate can also be reduced ($<0.5\%$) and finally by implementing AR coatings Fresnel reflections can be eliminated as well. With optimisation the throughputs of each sub-aperture of the rampper are projected to be $>68\%$. If we assume all other non-metallic optics within the IPRI are AR coated as well, we project a total sub-aperture throughput through the entire IPRI of $>50\%$ should be achievable. 

\subsection{The Integrated Pupil-Remapping Interferometer}
\begin{figure}
\centering 
\includegraphics[width=0.90\linewidth]{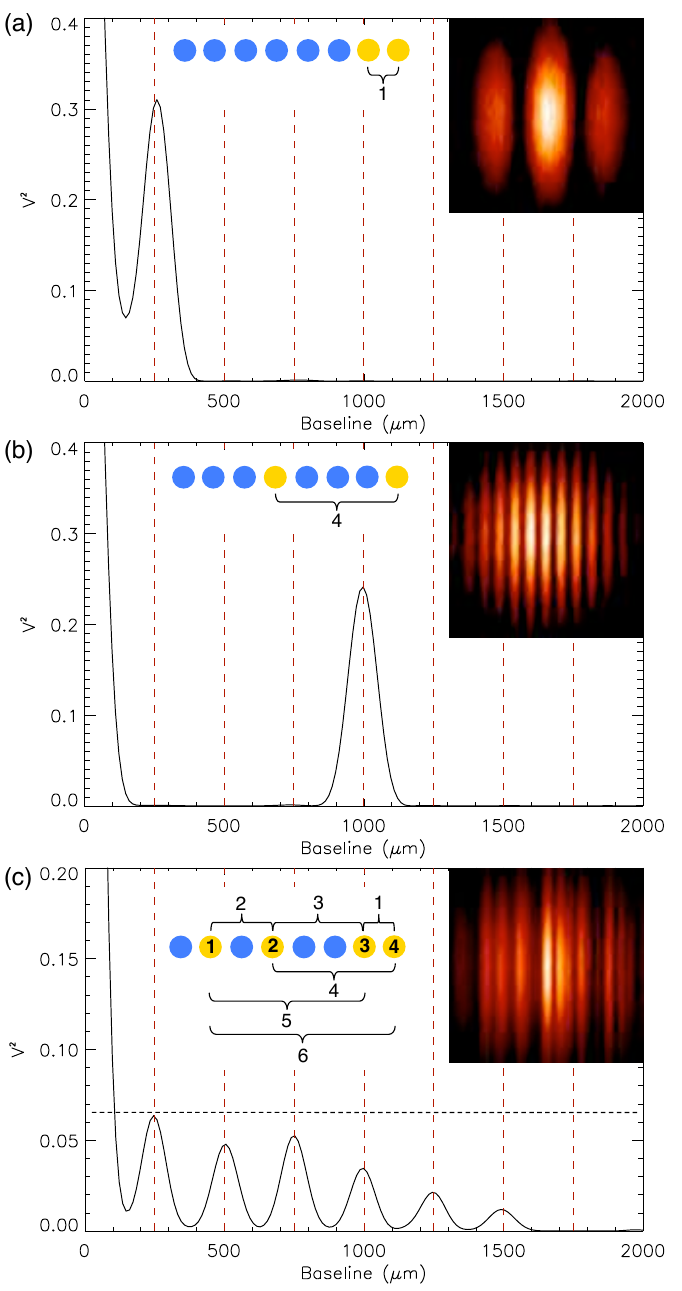}
\caption{Power spectra displaying the detected Fourier components for the corresponding set of waveguides that were used (highlighted by yellow dots while the unused waveguides are shown in blue). The insets show the recorded interference pattern in each case. The numbered parentheses indicate the baseline order with the shortest unit baseline corresponding to $250~\mu$m (the separation between neighbouring waveguides).}
\label{fig:nrfringes}
\end{figure}

Prior to commissioning at the telescope, the pupil-remapper was aligned within the IPRI and tested in a laboratory environment with a $45$~nm bandwidth light source. The laboratory tests were conducted without the cross-dispersing prism and anamorphic optics. Initially, the $7$ possible interferometer baseline lengths were tested in turn, by misaligning the unwanted elements of the segmented mirror leaving only a pair of waveguides illuminated in each case. High visibility fringes were observed on all interferometer baselines, two of which are shown in the insets to Figure~\ref{fig:nrfringes}(a) and (b). The fringe pattern was windowed using a Hanning window and then integrated in the vertical direction in order to collapse each $2$-dimensional image into a $1$-dimensional line profile. A Fourier transform was calculated for the line profile, the square of which yielded the power spectrum and squared visibilities ($V^{2}$) of the pattern as shown in Figure~\ref{fig:nrfringes}(a) and (b). The average amplitude fluctuation of the visibilities of the $7$ interferometer baselines as a function of time was $2.2\%$ RMS.

Because the waveguides were equally spaced at the output of the pupil-remapper, a result of the fact that the remapper was designed to be used in tandem with a beam combining chip which takes this input format, a non-redundant subset was selected by the simple procedure of misaligning unwanted channels with the segmented mirror array. An interferometer with $6$ unique baselines was realised in this fashion using $4$ waveguides, of which the associated fringe pattern, power spectrum and squared visibilities are shown in Figure~\ref{fig:nrfringes}(c). The theoretical maximum squared visibility is proportional to $1/N^{2}$ where $N$ is the number of waveguides used (the basic mathematical foundations of interferometry are described in~\cite{Lawson2000} and~\cite{Monnier2003}); here this is $V^{2} = 0.063$ (assuming even illumination) and is marked by a dashed line in Figure~\ref{fig:nrfringes}(c). It can be seen that the instrument generated high visibility fringes on all $6$ baselines simultaneously which approach this value. The departure from the theoretical value for longer baselines can be attributed to aberrations in the ad-hoc optical setup downstream of the pupil remapper used to perform this particular test which caused the higher fringe frequencies to wash out, and was not a problem for our on-sky testing which used the far superior optical setup shown in Fig.~\ref{fig:dragonfly}. Nonetheless the high visibilities observed confirm that the arms of the instrument were matched to significantly better than the coherence length of the light ($\lambda^{2}/\Delta \lambda\approx54~\mu$m for the $45$~nm bandwidth light source used).

Owing to its resilience against atmospheric and instrumental phase noise, the closure phase has been the key observable for the detection of high contrast companions with conventional aperture masking~\citep{Lacour2011}. The closure phase is defined as the sum of the phase measured around a closing triangle of baselines in the spatial frequency domain. A closing triangle comes about because each aperture of the interferometer is responsible for 2 baselines and hence spatial frequency components and can be formed by using three sub-apertures, for example. It is insensitive to phase errors on individual channels, thus delivering an observable that is a function of the target object's intensity distribution only~\citep{Baldwin1986}. The stability of the closure phase as a function of time is of great importance as it is one of the parameters (in addition to the Fourier coverage and others) that define the maximum contrast of a detectable companion~\citep{Lacour2011}. Stabilities at or below the $1^{\circ}¡$ level will see the IPRI reach scientific utility, while closure phases $<0.5^{\circ}$ place the instrument on par with the present state-of-the-art achieved with aperture masking in conjunction with AO systems~\citep{Lacour2011,Kraus2012}. The IPRI's closure phase stability was tested with artificially induced phase offsets (via pistoning of the segmented mirrorÕs elements). It should be made clear that the pistoning was carried out with a single unidirectional sweep of the segmented mirror's elements, across their entire range, over a timescale of half-a-minute. This did not simulate the bidirectional, fast time scale phase fluctuations induced by the atmosphere. A single frame was collected for each position of the mirrors' elements during the sweep. The visibilities were calculated as outlined previously. The closure phase was calculated for each frame by multiplying the complex visibilities of $2$ of the Fourier components by the complex conjugate of the $3^{rd}$ that formed the closing triangle and then taking the argument of this value. The RMS fluctuation in the closure phase was $0.4^{\circ}$ over the piston scan. As the segments were pistoned one-at-a-time over a range which corresponded to several multiples of the wavelength of the light, then the $0.4^{\circ}$ RMS closure phase stability was measured with effective wavefront errors $>2\pi$. However, current AO systems can deliver residual wavefront errors of $\approx250$~nm while extreme AO systems aim for $\approx80$~nm (SPHERE \citep{Beuzit2008} performance for median seeing privately communicated by Boccaletti)~\citep{Bouchez2009}. Approximating the scaling of phase errors to be linear in the case of well-corrected wavefronts \citep{Martinache2010} yields expected closure phase stabilities of $0.18^{\circ}$ RMS and $0.058^{\circ}$¡ RMS when used in tandem with an AO and an extreme AO system respectively (where we have also accounted for the effect of a random error on every sub-aperture simultaneously). The closure phase stability projected for use with an AO system is a factor of $3$ better than the $0.5^{\circ}$ threshold at which the IPRI becomes scientifically competitive while for extreme AO systems it is almost an order of magnitude below the state-of-the-art achieved in masking. We do not claim that this simple analysis can directly translate to the predicted on-sky performance where several sources of additional systematic error may arise. However this promising level of stability suggests that the instrument has the potential to compete with aperture masking as will be discussed in greater detail in the later sections.

\subsection{On-sky observations}
The IPRI was tested on-sky on the nights of the $20^{th}$--$21^{st}$ of May, $2011$ on the $3.9$-m Anglo-Australian Telescope (AAT) at Siding Spring Observatory. As this telescope was not equipped with AO, the injection into the waveguides was severely degraded and interference fringes were in constant motion due to the turbulent phase perturbations imposed by the atmosphere. The median seeing for Siding Springs in May was $1.8$~arcsec at $0.55~\mu$m (from \textit{aao.gov.au}) which resulted in a Fried parameter, $r_{0}=6.3$~cm and more relevantly $r_{0}=22$~cm at $1.55~\mu$m.This value of $r_{0}$ was $2.5\times$ smaller than the projection of each micro-lens on the primary mirror ($56$~cm) and hence the atmospherically induced wavefront errors were $>\pi$~radians across each micro-lens. This resulted in poor coupling of the starlight into the waveguides.

\begin{figure}
\centering 
\includegraphics[width=0.99\linewidth]{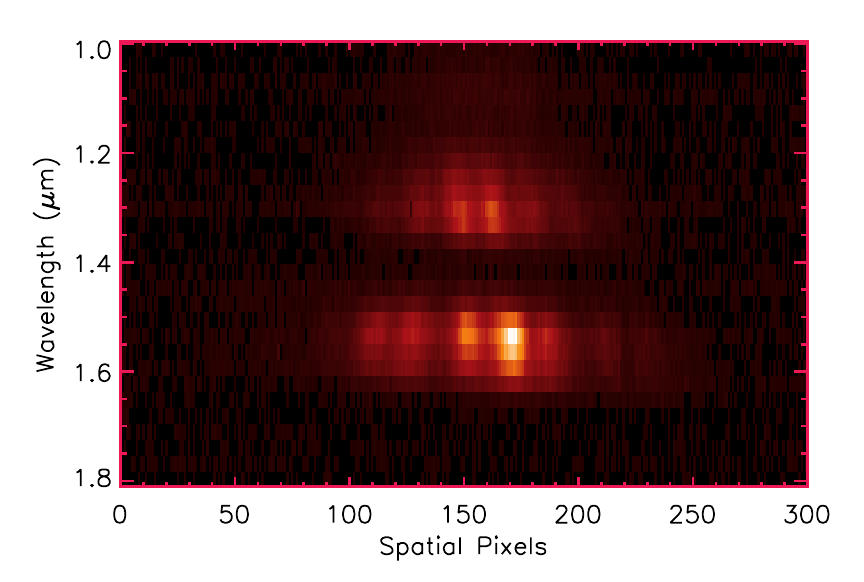}
\caption{A single image of a spectrally dispersed fringe pattern taken on-sky at the AAT while observing Antares.}
\label{fig:fringes}
\end{figure}

In order to minimise the loss of wavefront coherence, short integration times were necessary but were limited by the readout noise of the InGaAs camera used ($200$ electrons while a scientifically competitive HAWAII-$2$RG is $<18$ electrons). As a compromise our observations were taken with $200$~ms exposures, which were many times longer than the typical atmospheric coherence time ($\tau_{0}\approx30$~ms at $1.55~\mu$m for Siding Springs). The target object was the red supergiant Antares, one of the brightest objects in the near-infrared sky which was also at low air mass. A data set of $1000$ consecutive frames was collected. An example single exposure (after cleaning and flat-fielding) extracted from the data series is depicted in Figure~\ref{fig:fringes}. Clear fringes corresponding to the strongest baseline can be seen, with the slight tilt from vertical exhibited by the fringes with wavelength implying that the path-lengths through the interferometer generating that particular fringe were temporarily unbalanced (due to the varying atmospheric column depths above each sub-pupil). 

\begin{figure}
\centering 
\includegraphics[width=0.93\linewidth]{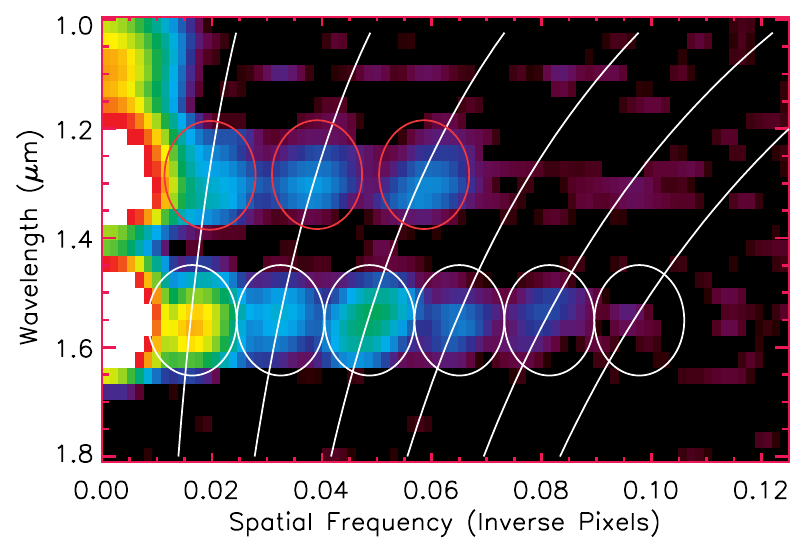}
\caption{Power spectrum of the interference fringes recorded by the instrument. The horizontal axis gives the spatial frequency, in inverse pixels, for the Fourier components. All fringes were well sampled. The white lines depict the theoretically expected spatial frequencies of fringes for our given remapping chip, reimaging optics and computed dispersion of the prism. The red and white circles highlight the detected spatial frequency components in J and H bands respectively.}
\label{fig:powerspec}
\end{figure}

The cumulative power spectrum shown in Figure~\ref{fig:powerspec} was obtained by summing the squared one-dimensional Fourier transform over the $1000$ frames. Two wavelength bands are clearly visible around $1.55~\mu$m (H-band) and $1.3~\mu$m (J-band). Six peaks in the power spectrum can be seen in the H-band (marked by white circles), which show that fringes were recorded on all $6$ baselines of the instrument. The system visibilities for these channels were calibrated by removing the effect of unequal beam fluxes (which resulted from the unequal throughputs of the waveguides) in post processing, yielding a mean visibility of $0.31$.    As no photometric signals were used to monitor the coupling into each waveguide, the rescaling was done based on the laboratory measurements of the throughputs, which was a reasonable first-order assumption. It should be made clear that the on-sky visibilities have been calibrated so that the maximum value each baseline can take is $1$, which is in contrast to the laboratory visibilities which were not calibrated. Up to $15\%$ of the reduction in visibility can be attributed to the fact that Antares was partially resolved by the AAT ($41.3$ mas angular size~\citep{Richichi1990}) leading to a recalibrated mean system visibility of $0.36$ for the longest baseline. 

A full numerical model of the telescope-atmosphere system was constructed to interpret the commissioning data. The system simulator included a single-layer phase screen with random fluctuations statistically matching a Kolmogorov turbulent spectrum. Adopting median Siding Springs values for seeing and windspeed ($7$~m/sec \textit{aao.gov.au}) in May, the code produced cross-dispersed interferograms. Light losses, beam asymmetries, photon/readout noise, and the relatively long integration time of $200$~ms were all incorporated in the model. Results from the reduction of the simulated data gave median system visibilities of $0.39$ in the H-band, which were consistent with the on-sky value of $0.36$ for the longest baseline.The visibilities had a signal-to-noise of about one from a single frame of data (both in simulations and as observed in the on-sky data), which is the theoretically expected value when imaging in the speckle regime~\citep{Roddier1988} as was the case for our experiment due to the relatively long integration times. The simulator revealed that the majority of the degradation of the visibilities could also be attributed to temporal effects, where fringes were blurred due to the exposure time being significantly longer than the coherence time.

It can be seen that several spatial frequency peaks were also recovered in the J-band (marked by red circles). Although our instrument was not specifically designed for operation here, the short single-mode cutoff of the waveguides and the recovery of these spatial frequencies suggests a straightforward extension to the J-band in the future.

Using the same code it was also possible to compare the measured throughput with expectations, with the finding that detected counts were about a factor of $2$--$3$ lower than predicted. Such losses could easily arise if the seeing was a little worse than simulated, and/or from known issues with the microlens alignment with respect to the chip. Recovering expected throughputs is a key component in pushing the science towards a fainter class of scientifically compelling targets. Although these results indicate there is still some work to do, they are encouraging enough to project that high efficiency operation should be possible with a more mature optical setup.

Closure phases were extracted corresponding to all of the $4$ possible closing-triangles and we discuss the stability of the one with the highest signal-to-noise of these. Each of the $5$ spectral channels spanning the H-band yielded closure phases of $0\pm5^{\circ}$ standard error (zero is the expected closure phase for a nearly unresolved centrosymmetric object such as Antares), which were consistent with numerical simulations ($0\pm4^{\circ}$ standard error) that included seeing. We therefore conclude that the on-sky closure phase performance can be entirely attributed to uncorrected seeing (temporal fringe blurring) and hence there were no unexpected or unexplained drops in performance at the telescope as compared to the laboratory.

\section{Predicted on-sky performance with adaptive optics}\label{s:pred}
With the closure phase stability for the on-sky tests explained by the presence of seeing, it is reasonable to use the closure phase stability of the laboratory measurements as a proxy for predicting the on-sky performance of the instrument if used in conjunction with AO systems. It is important to understand that the following extrapolation of on-sky performance when used in conjunction with AO does not take into account all the systematic errors that are present when operating on a telescope and is meant only as a guide to the ball park performance that may be achieved in an ideal case scenario where the systematics can all be calibrated out. 

We can arrive at a predicted maximum contrast of detecting a companion as follows. As outlined previously, the laboratory measurement of the closure phase stability can be rescaled to account for wavefront errors across all sub-apertures simultaneously around a closing triangle with the magnitude equal to that expected of the residuals from an AO or an extreme AO system ($\approx250$ and $\approx80$~nm~\citep{Bouchez2009} respectively). This yields a $0.18^{\circ}$ RMS and $0.058^{\circ}$ RMS closure phase excursion, respectively. This level of closure phase stability can be converted into a limit for which a companion could be detected at an angular separation of $1\lambda/D$ from the stellar host (parameterized for a $7$-aperture system using equation $2$ in~\citet{Lacour2011}) to give a contrast of $\approx2200:1$ ($1\sigma$) for an AO system and $\approx7000:1$ ($1\sigma$) for an extreme AO system. When converted to contrasts for typical confidence levels ($3.3\sigma$, $99.9\%$) used in aperture masking observations we get $\approx750:1$ for AO and $\approx2100:1$ for extreme AO systems. For clarity this means that a companion which is $750$ times fainter than the parent star could be detected with a signal-to-noise of $3.3$ when used in conjunction with an AO system. These values compare favourably with the current state-of-the-art contrasts for aperture masking that have been achieved; $100$-$400:1$~\citep{Huelamo2011,Kraus2012}, but cannot be directly compared to the aperture masking results as they were captured in longer wavelength bands. The predicted levels of contrast would currently limit the H-band version of the instrument to exploring the Brown dwarf regime. However, by moving further into the mid-infrared (L'-band) where the stars are fainter and the young planets are therefore relatively brighter, Jupiter-like planets may potentially be within reach.

In practice, extrapolations of performance from the laboratory to the telescope are fraught with difficulty. For a start our above prediction might be regarded as pessimistic, for it strictly only applies to a small number of detected frames and random errors in the closure phase noise could be reduced by taking a longer data run (there is a limit to this as well). However, an important source of systematic error not considered thus far is that encountered when slewing between a science target and a point-spread-function reference star, and there may be persistent-speckle or non-common-path errors that come into play which degrade the maximum contrast achievable (although it is worth noting that extreme AO systems are specifically engineered to avoid such issues). It is also important to bear in mind that the system commissioned represents a very simple testbed; a final scientific instrument would benefit from two major scalings in data quantity (and quality). For simplicity of analysis, the closure phase statistics presented so far strictly apply to a single wavelength channel. Critical analysis of the ensemble of wavelength channels across the H-band not only boosts the SNR with the square root of sample size, but also delivers differential phase data. Zimmerman \textit{et al.} have shown a factor of $3$--$4$ improvement in SNR by exploiting such spectro-spatial data~\citep{Zimmermann2011}. The second major scaling is associated with the expansion of the array with a next-generation remapper capable of utilizing a larger fraction of the full pupil. For example a scheme which recombined $4$ sets of $7$ sub-apertures in non-redundant configurations which would be relatively simple to implement would yield $60$ independent closure phase measurements; a factor of two gain in contrast over the numbers quoted above. As our $\approx2000:1$ projection is already competitive with the state-of-the-art over this range of spatial scales, we will (conservatively) adopt this value in the discussion of scientific reach which follows, but note that there are substantial grounds for the belief that a next generation instrument could do significantly better.  

\section{Defining the science discovery space for an Integrated Pupil Remapping Interferometer}\label{s:space}
Three parameters are required to define the niche parameter space for exoplanets with the IPRI: the angular separation (between the planet and the star), the brightness of the star and the contrast between the planet and the star. The real strength of the IPRI lies in being able to detect companions at spatial scales of half to a few $\lambda/D$ from the parent star with moderate contrast ($\approx2000:1$ at $3.3\sigma$ as described above). In comparison, the Gemini Planet Imager (GPI) instrument which makes use of an extreme AO system and a Lyot coronagraph will target detections at $>3\lambda/D$~\citep{Macintosh2006}. There are several sophisticated coronagraph designs such as the phase-induced amplitude apodisation (PIAA) coronagraph being employed in the SCExAO instrument which aim to reduce this inner blind spot to $\approx1\lambda/D$~\citep{Lozi2009}. Although contrasts as high as $10^{6}:1$ have been demonstrated in a laboratory environment, wavefront calibration at separations around $1 \lambda/D$ have proven extremely difficult thus far and contrasts have not exceeded $100:1$ in this region to date. This means that our IPRI offers highly competitive contrasts for planet detection in the $1\lambda/D$ separation region for the immediate future against extreme AO/coronograph technologies. This is an extremely important region of the angular separation space when we consider the nearest star forming regions (Scorpius-Centaurus at $380-470$~lyrs and Taurus-Ariga $460$~lyrs). The solar-like stars in these regions are believed to harbour many young planets which are still warm and hence very luminous from the formation process making them easier to detect. An angular separation of $1\lambda/D$ for an $8$-m telescope in the H-band corresponds to $7$~AU (the approximate distance of Jupiter from the sun) at the distance to these star forming regions. This represents a sought-after parameter space as it corresponds to solar system scales ($\approx10$~AU from the parent star) and hence this forms the basis for the demand for instruments addressing the regime of small angular separations from the star.

At present, aperture masking would represent the chief competitor to the IPRI at small angular separations. Although the predicted contrast for the IPRI is superior to masking when used with AO, the current throughputs are lower on an aperture-by-aperture basis. This, as outlined in the experimental section above, we believe we can address in future to realise a remapping interferometer with $>50\%$ throughput across each sub-aperture. However, when using the IPRI in tandem with a next generation extreme AO system, it is not the throughput that will determine the limiting brightness of the host star (magnitude) that can be studied, it is the level of signal required for the wavefront sensor of the AO system to lock successfully. Once the AO system is locked, long integration times can be used to maximise the signal-to-noise ratio of the Fourier observables (only limited by readout noise). In this way the IPRI will extend even the reach of aperture masking in the region of small angular separations from the host star. 

In future the PIAA coronagraph being developed at Subaru telescope will also become a strong competitor in this angular separation regime. Although it currently offers throughputs $>50\%$ for separations $>2.2\lambda/D$ from the star~\citep{Lozi2009}, with new PIAA lenses it is projected to push this inner working angle to $0.6\lambda/D$. As the only other next generation high contrast imaging instrument being currently developed to image as close in as the IPRI to the parent star (ignoring aperture masking for the moment), it is interesting to note that the two instruments share a similar level of cost and complexity; the IPRI has a custom made photonic chip, a segmented mirror and $2$ micro-lens arrays while the PIAA has two sets of custom made lenses, a spider removal plate and a binary mask. This reinforces the fact that the IPRI is uniquely positioned in the field of high contrast imaging with the few competitors required to exploit similar levels of sophistication to compete. 

The pupil remapping technology pioneered here also represents a basic enabling platform for future photonic instrumentation. With it, light from a telescope pupil may be injected into a planar waveguide structure while preserving the inherent wavefront coherence. Harnessing the myriad of devices now in use from the photonics and communications industry for the manipulation of light in integrated chips, elements such as splitters, couplers, gratings and delay lines may all be incorporated into future instruments. One example of the use of such elements might be in the construction of a nulling interferometer, an instrument designed to null out the bright star with high suppression leaving behind the signal from the faint companion. This interferometric analogue of a coronagraph would yield still greater advances into high contrast imaging.

\section{Conclusion}
We present the first experimental results from an integrated pupil-remapping interferometer from both laboratory and on-sky testing. The powerful combination of micro-scale optical technologies, namely the segmented mirror, microlens arrays and the pupil-remapping chip, enables advanced photonic processing and control of starlight in a simple and elegant fashion. The instrument has demonstrated high visibility fringes and stable closure phases ($0.4^{\circ}$ RMS for $>2\pi$ wavefront error) from the first accurately path-length-matched, 3D photonic device consisting of uniquely routed waveguides in a laboratory setting. The on-sky tests showed performance metrics (throughputs, visibility and closure phase recovery) consistent with those expected given wavefronts corrupted by uncorrected atmospheric turbulence. We believe that with the use of extreme AO on larger telescopes both high visibilities and high levels of closure phase stability will be achieved. For a modest $7$-aperture system, the ball-park projection of the faint companion contrast attainable is $\approx2000:1$ ($3.3\sigma$) for an angular separation down to $1\lambda/D$ from the parent star. A key feature of our architecture is the ease with which it may be dramatically scaled by populating the entire pupil plane with waveguides. An IPRI offering near-complete Fourier coverage and spectrally dispersed data, coupled with predicted performance levels of closure phase stability from an extreme AO equipped telescope would deliver unsurpassed performance over the most critical region of angular separation: that corresponding to solar system scales in the nearest star-forming regions. The possibility of such exciting performance illustrates the immense promise for new-generation astrophotonic instrumentation to make unique contributions to observational astrophysics.

\section{Acknowledgements}
The authors would like to acknowledge the efforts of Macquarie Engineering and Technical Services (METS) in getting the Dragonfly instrument ready for the telescope tests. The authors would like to also acknowledge Conor Harll for producing figure \ref{fig:dragonfly}. This research was conducted with the Australian Research Council Centre of Excellence for Ultrahigh Bandwidth Devices for Optical Systems (project number CE$110001018$) and the assistance of the LIEF and Discovery Project programs. This work was supported by the OptoFab node of the Australian National Fabrication Facility. We also thank the operations staff at the Anglo-Australian Telescope for providing invaluable assistance during the telescope tests, and the AAO Director, Matthew Colless, for providing telescope time for this experimental program.

\label{lastpage}

\begin{thebibliography}{}
   \bibitem[\protect\citeauthoryear{Angel \& Woolf}{1997}]{Angel1997}                    					
  Angel J. R. P., Woolf N. J., $1997$, ApJ, $475$, $373$
  
  \bibitem[\protect\citeauthoryear{Baldwin}{1986}]{Baldwin1986}                    					
  Baldwin J. E., Haniff C. A., Mackay C. D., Warneret P. J., $1986$, Nature, $320$, $595$
  
  \bibitem[\protect\citeauthoryear{Benisty et al.}{2009}]{Benisty2009}                    				
  Benisty M., Berger J.-P., Jocou L., Labeye P., Malbet F., Perraut K., Kernet P., $2009$, A\&A, $498$, $601$
  
  \bibitem[\protect\citeauthoryear{Berger et al.}{2001}]{Berger2001}                      				
  Berger J. P., et al., $2001$, A\&A, $376$, L$31$ 
  
  \bibitem[\protect\citeauthoryear{Beuzit et al.}{2008}]{Beuzit2008}         
  Beuzit J.-L., et al., $2008$, Proceedings of SPIE, $7014$, $701418$
  
  
  \bibitem[\protect\citeauthoryear{Bouchez et al.}{2009}]{Bouchez2009}                    			
  Bouchez A. H., et al., $2009$, Proceedings of SPIE, $7439$, $74390$H  
         
  \bibitem[\protect\citeauthoryear{Bracewell}{1978}]{Bracewell1978}      
 Bracewell, R. N. 1978, Nature, 274, 780
  
  \bibitem[\protect\citeauthoryear{Burocki et al.}{2010}]{Burocki2010}                    				
  Burocki W. J., et al. 2010, Science, 327, 977
  
  \bibitem[\protect\citeauthoryear{Chang et al.}{1998}]{Chang1998}     
  Chang M.~P., Buscher D.~F., $1998$, Proceedings of SPIE, $3350$--$2$
  
  \bibitem[\protect\citeauthoryear{Coude du Foresto}{1994}]{Coude1994}                    			
  Coude du Foresto V., 1994,  Proceedings of the $158th$ IAU, $261$
  
   \bibitem[\protect\citeauthoryear{Charles et al.}{2012}]{Charles2012}                    					
  Charles N., et al., $2012$, Appl. Opt., 21, Accepted on the $10^{th}$ of September $2012$.
  
  \bibitem[\protect\citeauthoryear{Eaton, Zhang \& Herman}{2005}]{Eaton2005}                   		
  Eaton S. M., Zhang H., Herman P., $2005$, Opt. Express, $13$, $4708$ 
  
  \bibitem[\protect\citeauthoryear{Gattass \& Mazur}{2008}]{Gattass2008}                    			
  Gattass R. R., Mazur E., 2008, Nat. Photonics, $2$, $219$
  
  \bibitem[\protect\citeauthoryear{Guyon}{2003}]{Guyon2003}                    					
  Guyon O., 2003, A\&A, $404$, $379$
  
  \bibitem[\protect\citeauthoryear{Huby et al.}{2012}]{Huby2012}        
  Huby E., et al., 2012, A\&A, $541$, A$55$
  
  \bibitem[\protect\citeauthoryear{Huelamo et al.}{2011}]{Huelamo2011}                    			
  Huelamo N., Lacour S., Tuthill P., Ireland M., Kraus A., Chauvin G., $2011$, A\&A, $528$, L$7$
  
  \bibitem[\protect\citeauthoryear{Jovanovic et al.}{2012}]{Jovanovic2012}    
  Jovanovic N., Spaleniak I., Gross S., Ireland M., Lawrence J., Miese C., Fuerbach A., Withford M., $2012$, Opt. Express, $20$, $17029$
  
  \bibitem[\protect\citeauthoryear{Kotani et al.}{2010}]{Kotani2010}							
  Kotani T., et al., Lecavelier des Etangs, A., Vidal-Madjar, A., $2010$, Proceedings of SPIE, $7734$, O$1$ 
  
  \bibitem[\protect\citeauthoryear{Kraus \& Ireland}{2012}]{Kraus2012}                     			
  Kraus A., Ireland M. J., $2012$, ApJ, $745$, $1$
 
  \bibitem[\protect\citeauthoryear{Kraus et al.}{2008}]{Kraus2008}                     				
  Kraus, A., Ireland, M. J., Martinache, F., Lloyd, J. P., $2008$, ApJ, $679$, $762$
  
  \bibitem[\protect\citeauthoryear{Kraus et al.}{2011}]{Kraus2011}                    			
  Kraus A., Ireland M. J., Martinache F., Hillenbrand L. A., $2011$, ApJ, $731$, $8$
 
  \bibitem[\protect\citeauthoryear{Labadie et al.}{2007}]{Labadie2007}    
 Labadie L., Le Coarer E., Maurand R., Labeye P., Kern P., Arezki B., Broquin J.-E., $2007$, A\&A, $471$, $355$
 
  \bibitem[\protect\citeauthoryear{Lacour, Thiebaut \& Perrin}{2007}]{Lacour2007}                      	
  Lacour S., Thiebaut E. J., Perrin G., $2007$, MNRAS, $374$, $832$ 
 
  \bibitem[\protect\citeauthoryear{Lacour et al.}{2011}]{Lacour2011}                    				
  Lacour S., Tuthill P., Amico P., Ireland M., Ehrenreich D., Huelamo N., Lagrangeet A.-M., $2011$, A\&A, $532$, $72$
  
   \bibitem[\protect\citeauthoryear{Lang et al.}{1994}]{Lang1994}   
  Lang, T., Thevenaz, L., Ren, Z., B., Robert, P., $1994$, Meas. Sci. Technol., $5$, $1124$
  
  \bibitem[\protect\citeauthoryear{Laurent et al.}{2002}]{Laurent2002}                     				
  Laurent E., et al., $2002$, A\&A, $390$, $1171$ 
 
 \bibitem[\protect\citeauthoryear{Lawson}{2000}]{Lawson2000}   
 Lawson P. R., editor, $2000$, \textit{Principles of Long Baseline Stellar Interferometry}, $1999$ Michelson Summer School
 
  \bibitem[\protect\citeauthoryear{Lloyd et al.}{2006}]{Lloyd2006}                  					
  Lloyd J., Martinache F., Ireland M. J., Monnier J. D., Pravdo S. H., Shaklan S. B., Tuthill P. G., $2006$, ApJ, $659$, L$131$
  
  \bibitem[\protect\citeauthoryear{Lozi, Martinache \& and Guyon}{2009}]{Lozi2009}                    	
  Lozi J., Martinache F., Guyon O., $2009$, PASP, $121$, $1232$  
 
  \bibitem[\protect\citeauthoryear{Macintosh et al.}{2006}]{Macintosh2006}                   			
  Macintosh B., et al., Proceedings of SPIE, $672$, $62720$L 
  
  \bibitem[\protect\citeauthoryear{Marois et al.}{2010}]{Marois2010}                    				
  Marois C., Zuckerman B., Konopacky Q. M., Macintosh B., Barman T., $2010$, Nature, $468$, $1080$
  
  \bibitem[\protect\citeauthoryear{Martinache}{2010}]{Martinache2010}   
  Martinache F., $2010$, ApJ, $724$, $464$ 
  
  \bibitem[\protect\citeauthoryear{Monnier}{2003}]{Monnier2003}   
 Monnier J. D., $2003$, Reports on Progress in Physics, $66$, $789$
  
  \bibitem[\protect\citeauthoryear{Nolte et al.}{2003}]{Nolte2003}                     					
  Nolte S., Will M., Burghoff J., Tuennermann A., $2003$, Appl. Phys. A, $77$, $109$ 
 
  \bibitem[\protect\citeauthoryear{Norris et al.}{2011}]{Norris2011}  
  Norris B., et al., $2011$, Proceedings of IQEC/CLEO Pac-Rim, $5410$-CT-$4$ 
 
  \bibitem[\protect\citeauthoryear{Perrin et al.}{2006}]{Perrin2006}                     				
  Perrin G., Lacour S., Woillez J., Thiebaut E. J., $2006$, MNRAS, $373$, $747$ 
 
  \bibitem[\protect\citeauthoryear{Richichi \& Lisi}{1990}]{Richichi1990}                    				
  Richichi A., and Lisi F.,  $1990$, A\&A, $230$, $355$ 
 
 \bibitem[\protect\citeauthoryear{Roddier}{1988}]{Roddier1988} 
  Roddier F., $1988$, Physics Reports, $170$, $97$

  \bibitem[\protect\citeauthoryear{Serabyn, Mawet \& Burruss}{2010}]{Serabyn2010}                      
   Serabyn E., Mawet D., Burruss R., $2010$, Nature, $464$, $1018$
 
 \bibitem[\protect\citeauthoryear{Shaklan \& Roddier}{1988}]{Shaklan1988}    
  Shaklan S., Roddier F., $1988$, Appl. Opt., $27$, $2334$
   
  \bibitem[\protect\citeauthoryear{Thomson et al.}{2011}]{Thomson2011}                    			
  Thomson R. R., Birks T. A., Leon-Saval S.G., Kar A. K., Bland-Hawthorn J., $2011$, Opt. Express, $19$, $5698$ 
  
  \bibitem[\protect\citeauthoryear{Tinney et al.}{2002}]{Tinney2002}                    				
  Tinney C. G., Butler R. P., Marcy G. W., Jones H. R. A., Penny A. J.,  McCarthy C., Carteret B. D., $2002$, ApJ, $571$, $528$
  
  \bibitem[\protect\citeauthoryear{Tuthill et al.}{2000}]{Tuthill2000}                    				
  Tuthill P., Monnier J. D., Danchi W. C., Wishnow E. H., Haniff C. A., $2000$, PASP, $112$, $555$
  
  \bibitem[\protect\citeauthoryear{Tuthill et al.}{2010}]{Tuthill2010}                      				
  Tuthill P. et al., $2010$, Proceedings of SPIE, $7734$--$59$ 
 
  
   \bibitem[\protect\citeauthoryear{Zimmermann et al.}{2011}]{Zimmermann2011}                    		
   Zimmermann N., et al., $2011$,  Bulletin of the American Astronomical Society, $43$ 
   
\end{thebibliography}
\end{document}